\documentclass[aps,showpacs,twocolumn,longbibliography,nofootinbib,secnumarabic]{revtex4-1}
\usepackage{graphicx}
\usepackage{dcolumn}
\usepackage{bm}
\usepackage{braket}
\usepackage{leftidx}
\usepackage{cancel}
\usepackage{amsmath}
\usepackage{epstopdf}
\usepackage{color}
\usepackage[mathcal]{eucal}
\usepackage{float}

\usepackage{yfonts}
\usepackage{color}

\usepackage{scalerel}

\begin{document}
\date{\today}
\title{Entanglement, violation of Kramers-Kronig relation and curvature in spacetime}

\author{Mehmet Emre Tasgin}

\affiliation{Institute of Nuclear Sciences, Hacettepe University, 06800, Ankara, Turkey}
\affiliation{metasgin@hacettepe.edu.tr and metasgin@gmail.com}

\begin{abstract}
Recent studies show that a maximally entangled Schwinger pair creates a nontraversable Einstein-Rosen~(ER) bridge (wormhole) in the gravitational theory, which bridges causally not connected parts of the AdS spacetime~[PRL 111, 211603 and 211602]. Same authors raise the possibility of a general correspondence between a weaker entanglement and curvature. Here, we provide some clues for such a generalized relation. First, we show that (i) entanglement and (ii) violation of Kramers-Kronig~(KK) relations appear at the same critical parameters for a standard two-mode (squeezing) entanglement interaction. We also include the dampings. Second, we bring a study into attention: Presence of a spacetime-curvature in vacuum polarization, electron-positron Schwinger pairs, makes QED also violate the KK relations without violating the causality. Then, we discuss that these findings are possible to provide new clues for a generalized relation between entanglement and spacetime-curvature. Such an interpretation may also save violation of KK relations from implying the violation of the causality.
\end{abstract}

\maketitle

Entanglement poses a spooky action between a pair of well-separated particles. Although entanglement cannot be used for instant communication~\cite{controled_SL_communication}, choice of the measurement on one of the particles affect the state of the other instantaneously~\footnote{In a many-particle (ensemble) entangled state, measurement on a single particle can affect the state of all remaining particles. This appears in systems like single-photon superradiance or interacting Bose-Einstein condensates~\cite{tasgin2017many,sorensen2001Nature}.}. Similarly, a nontraverable wormhole, a phenomenon emerging in general relativity (gravitational theory), can connect (shortcut) two distant particles at a space-like separation. Although a nontraversable wormhole cannot be used for instant communication, as in entanglement, action on one of the particles can affect the state of the other one~\cite{Susskind2013cool}.

Recent influential works~\cite{sonner2013holographic,jensen2013holographic},\cite{Susskind2013cool} demonstrate a concrete relation between the two phenomena; entanglement in quantum theories and Einstein-Rosen~(ER) bridge (a wormhole) in the gravitational theory. In simple words, Refs.~\cite{sonner2013holographic,jensen2013holographic} use equivalence (holography) principle~\cite{maldacena1999large,witten1998anti} between the solutions of conformal (quantum) field theory~(CFT) and super-symmetric gravitational theory (as a low energy string theory), i.e. AdS${}_5$~\cite{fujita2011aspects}. On the CFT side, they provide analytical solutions~\cite{xiao2008exact} to a quark-antiquark pair, in a \textit{maximally entangled} state, created via Schwinger effect~\cite{semenoff2011holographic} in vacuum. The two particles are not in causal contact such that the light emitted from one particle cannot reach the other one~\cite{jensen2013holographic}. On the AdS side, they explicitly show that this maximally entangled state, of CFT, produces a nontraversable wormhole (ER bridge) in AdS~\cite{sonner2013holographic,jensen2013holographic,semenoff2011holographic}, via the mathematical correspondence (holography)~\cite{xiao2008exact} between 4-dimensional CFT and AdS${}_5$. A ``tunneling instanton'' solution~\footnote{A solution who can change (tunnel), e.g., between two minimum energies almost instantaneously~\cite{dunne2005worldline,dunne2006worldline}.} accompanies the Scwinger effect~\cite{semenoff2011holographic}, which corresponds to an ER bridge~\cite{kim1997classical,gutperle2002instantons,gorsky2002schwinger} between the horizons of the particles out of causal contact~\cite{jensen2013holographic,semenoff2011holographic}.

Such an explicit demonstration of the correspondence between a \textit{maximally entangled} Schwinger pair and emergence of a nontraversable ER bridge (wormhole) led Sonner~\cite{sonner2013holographic} raise the following intriguing question. What if the system is not in a maximally entangled state but in a \textit{weaker inseparable} state? 

In this paper, we provide new \textit{clues} on the entanglement-ER bridge correspondence. Our results show a possible extension of the connection between entanglement and ER bridge (very special solutions of Einstein Field equations, EFE, in curved spacetime) into a kind of correspondence between (i)~a particular, small, value of entanglement and (ii)~presence of curvature in spacetime.

We handle the correspondence in a completely different point of view. Instead of CFT, we use the standard second-quantized theory of quantum optics and QED in the curved spacetime~\cite{hollowood2008refractive,hollowood2007causality}. Similar to Refs.~\cite{sonner2013holographic,jensen2013holographic,semenoff2011holographic}, QED in curved spacetime deals with electron-positron Schwinger pairs, i.e. vacuum polarization. 

\begin{figure}
\begin{center}
\includegraphics[width=0.5\textwidth]{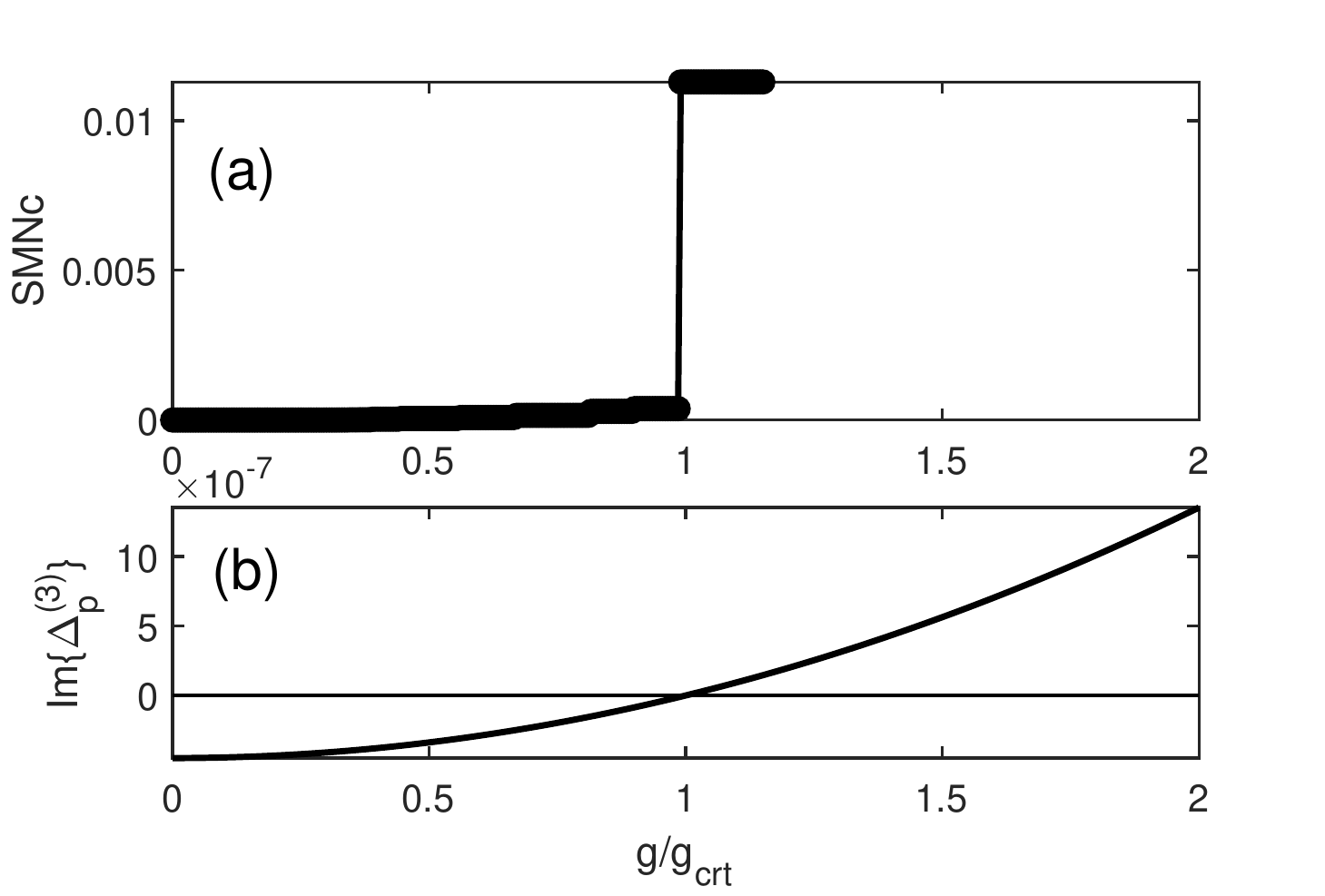}
\caption{For $\Delta=-\omega_m$, $\hbar g_1(\hat{c}^\dagger \hat{a}^\dagger + \hat{a}\hat{c})$ entangler interaction is favored. In this case, (a) single-mode nonclassicality~(SMNc) and (b) violation of KK relations appear at the same critical coupling $g$. In graphics (a), plot is ended at $g=1.2$. After that value system becomes unstable. }
\label{fig1}
\end{center}
\end{figure}

\begin{figure}
\begin{center}
\includegraphics[width=0.4\textwidth]{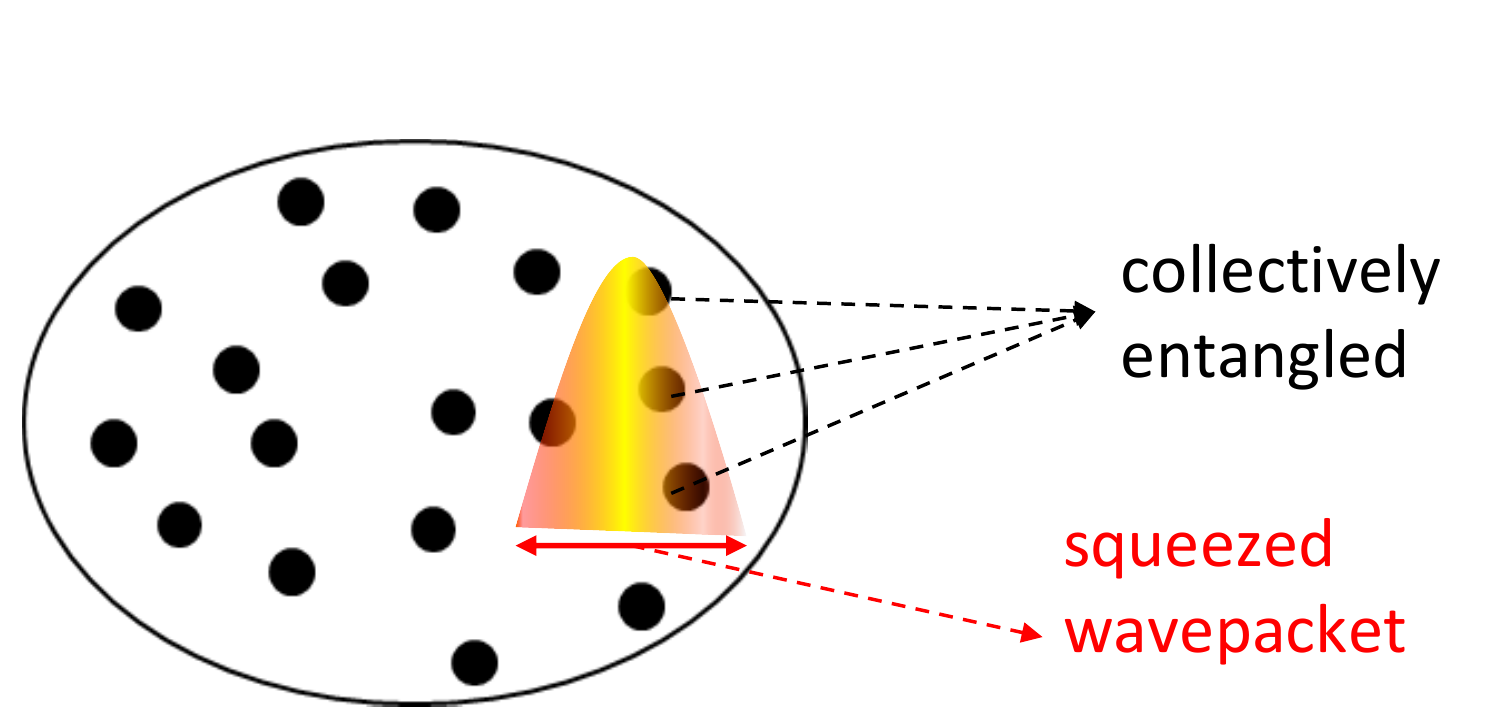}
\caption{If, for instance, a phonon is squeezed the the motional degree of freedom of the vibrating atoms are collectively entangled within the extent of the phonon wavepacket. A similar visualization can be made for a squeezed photon, e.g. the cavity mode.}
\label{fig2}
\end{center}
\end{figure}

First, we show that the most standard, exactly solvable, two-mode squeezing $\hat{\cal H}=\hbar (g_1\hat{c}^\dagger\hat{a}^\dagger+g_1^*\hat{a}\hat{c})$, interaction~\cite{josse2004entanglement} in a cavity manifests the violation of Kramers-Kronig~(KK) relations in the input/output (transfer functions) of the cavity.  More interestingly, we show that in such a damped cavity, nonclassicality~[entanglement or single-mode nonclassicality~(SMNc), e.g. squeezing] shows up at exactly the same critical (e.g. cavity-mirror) coupling, $g_1$, parameter. Moreover, this coincidence appears for any values of the system parameters! (see Fig.~\ref{fig1}) We note that SMNc of a (collective) quasi-excitation is the collective entanglement of the background (particles) generating the. e.g. squeezed, excitation~\cite{tasgin2017many,superfluid_spacetimePRL2014}, see Fig.~\ref{fig2} for an illustration.

Second, we examine (compare with) the quantum electrodynamics~(QED) solutions of vacuum-polarization in curved spacetime. Hollowood and Shore calculate the refractive index $n(\omega)$ for vacuum-polarization (Schwinger production of electron-positron pairs) of light in the presence of background curvature in spacetime~\cite{hollowood2008refractive,hollowood2007causality} using an unperturbed (sigma wordline~\cite{feynman1950mathematical,schwinger1951theory,bastianelli2002worldline,schubert2001perturbative}) formalism. They provide worldline instanton solutions describing Schwinger pair creation which can be interpreted as \textit{tunneling} in this formalism~\cite{hollowood2008refractive,hollowood2007causality}. They show that refractive index~\footnote{of the light which creates the vacuum polarization} violates the KK relations due to the interaction of electron-positron Schwinger pairs with a weak background curvature~\footnote{Here, light interacts with the background curvature through the created electron-positron pairs.}. We kindly remind that in Refs.~\cite{sonner2013holographic,jensen2013holographic,semenoff2011holographic}, maximally entangled Schwinger pairs are responsible for the ER bridge~(tunneling instanton).

Fortunately, Hollowood and Shore show that in the presence of background curvature in spacetime, violation of KK relations does not imply the violation of causality~\cite{hollowood2008refractive,hollowood2007causality,shore2003quantum}. Shortly, they argue that interaction of the generated electron-positron pair with the background ``curvature" violates only the strong equivalence principle~(SEP)~\cite{shore2003quantum,drummond1980qed} in general relativity~\footnote{SEP is not violated if the interaction depends only on the Christoffel symbols but not directly on the curvature~\cite{bertotti1990strong,shore2002faster,shore2003quantum}.}. Validity of SEP would imply the existence of global reference frames and this would indicate  a superluminal propagation which could violate the causality~\cite{shore2003quantum}. Weak equivalence principle, however, implies only the existence of local inertial reference frames which is not sufficient to establish a link between superluminal propagation and violation of causality~\cite{shore2003quantum}~\footnote{An extended discussion can be found in Refs.~\cite{shore2003quantum}}.

Both (i) a weakly-entangled system and (ii) a system, where a background spacetime curvature is present, violate the Kramers-Kronig relations. Hence, considering a possibility of a relation between entanglement and spacetime curvature could allow us to (a) circumvent the appearance of violation of KK relation to imply the causality violation in entangled devices and (b) extend the entanglement-ER bridge correspondence, bearing in Refs.~\cite{sonner2013holographic,jensen2013holographic,semenoff2011holographic}, into an entanglement-curvature duality. 

In advance, we state that our findings \textit{do not provide a proof} for the entanglement-curvature (spacetime) relation. The reason we raise the common appearance of violation of KK relations, showing up both in the onset of (i) weak entanglement and (ii) weak spacetime curvature, as a \textit{clue} is the following. Actually there are several reasons for this. First, KK violation is obviously not a common phenomenon which otherwise would not be so controversial to the audience. Second, Refs.~\cite{sonner2013holographic,jensen2013holographic,semenoff2011holographic} already provide a direct derivation between maximally entangled Schwinger pair and ER bridge. So, such a relation between weak entanglement and spacetime curvature is already something expected~\cite{sonner2013holographic}. Third, violation of KK and onset of entanglement appear at exactly the same critical coupling $g_1>g_{\rm gcr}$, see Fig.~\ref{fig1}. This obviously provides a stronger \textit{clue}, e.g. compared to Ref.~\cite{bruschi2016weight} where (fourth) small fluctuations in entanglement are shown to introduce curvature (weight) in a non-dissipative system. That is, there is no critical point in the valuable work~\cite{bruschi2016weight}, but the two critical values coincide in our case. Fifth, the observation on the violation of KK relations via entanglement or SMNc (especially in a continuous regime of variables, i.e. not at some resonances), ``forces'' us to believe in (check the existence of) such a correspondence. Induction of curvature with inseparability can avoid the violation of causality~\cite{hollowood2008refractive,hollowood2007causality,shore2003quantum} appearing via entanglement~\cite{sonner2013holographic}. 

So this work aims to gather all clues in a single box besides providing new ones. We propose a way for circumventing acceptance of the presence of the violation of causality under the light of Refs.~\cite{sonner2013holographic,jensen2013holographic,hollowood2008refractive,hollowood2007causality}. We underline that the work presented here has its roots at Ref.~\cite{tasgin2015mutual}.

{\bf \small Tunneling and KK relations}--- Before demonstrating our results in the next paragraphs, it is appropriate to discuss an important point. We note that violation of KK relations are observed also in ``causal'' interferometry devices~\footnote{M. Suhail Zubbairy --private communication (group meeting).}~\cite{beck1991group,wang2002causal,stern2012transmission} where multiple interferences take place. It is well-known that interference can avoid/limit electromagnetic field to occupy particular spatial domains. These domains are tunneled~\cite{chiao1999tunneling}. In these systems, the major problem is to define the ``tunneling times" for photons~\cite{davies2005quantum} which are calculated to be superluminal~\cite{wangPRA2007theoretical,winfulNature2003,winfulPRL2003}. Although this is discussed to appear due to the instantaneous spreading of the wavefunctions~\cite{hegerfeldt1998instantaneous,hegerfeldtPRD1974remark}, also relativistic equations demonstrate the superluminal tunneling times~\cite{perezPRD1977localization,hegerfeldtPRD1980remarks}. It is also shown that, still open to further questions, weak Einstein causality (which states that expectations or ensemble averages may not be superluminal) is not violated in tunneling process~\cite{wangPRA2007theoretical}. In Ref.~\cite{tasgin_group_index}, we discuss the interference phenomenon in details. In Ref.~\cite{tasgin_group_index}, we also show that the faster-than-light ``peak" velocities observed in various experiments~\cite{ChuPRL1982SL} are superluminal since the group-index (mathematically can be shown to govern the peak velocity~\cite{Jackson_book}) violates the KK relations. This provides a support for superluminal propagation needs to be accompanied by a violation of KK relations which Refs.~\cite{hollowood2008refractive,hollowood2007causality,shore2003quantum} show: does not necessarily imply the violation of causality. 

Actually the appearance of nonanalyticity in the upper-half of the complex-frequency-plane~(CFP), in our system, is different than the ones discussed in ``causal" interferometry devices~\cite{wang2002causal,Zubairy2014counterintuitive,wang2016counterintuitive}. The nonanalyticity we observe does not depend on the cavity length, unlike interferometry devices, hence possibly do not originate from interference\footnote{Actually it is not so important whether it appears via interference or not. Statement ``entanglement induces (or induced with) superluminal tunneling" also lines with our discussions, which strongly accompanies the tunneling instanton in Refs.~\cite{sonner2013holographic,jensen2013holographic,hollowood2008refractive,hollowood2007causality} wherein tunneling time can be more superluminal (faster) compared to Refs.~\cite{wangPRA2007theoretical,winfulNature2003,winfulPRL2003}.}. Our system is merely absorptive where violation of KK relations do not appear~\cite{Zubairy2014counterintuitive,wang2006superluminal}.

{\bf \small Entanglement \& violation of Kramers-Kronig relations}--- First, we consider an optomechanical system~\cite{genes&VitaliPRA2008robust,vitaliPRL2007optomechanical} in which a cavity mode $\hat{c}$ interacts with the vibrating mirror $\hat{a}$ placed inside the cavity. Then, we tune the cavity to favor $\hat{\cal H}_{\rm int}=\hbar (g_1\hat{c}^\dagger\hat{a}^\dagger + g_1^*\hat{a}\hat{c})$ two-mode squeezing~\cite{genes&VitaliPRA2008robust} type interaction. We show that entanglement and violation of KK relations appear at the same critical coupling $g=g_{\rm crt}$, in Fig.~\ref{fig1}.

Entanglement features of an optomechanical system have already been studied extensively~\cite{marquardt2009optomechanics}. Hamiltonian can be written~\cite{genes&VitaliPRA2008robust,tasgin2015mutual} as
\begin{eqnarray}
\hat{\cal H}= \hbar\Delta_c\hat{c}^\dagger\hat{c} + \hbar\omega_m\hat{a}^\dagger \hat{a} + \hbar g\hat{c}^\dagger\hat{c}\hat{q} 
+ i\hbar\varepsilon_{\rm L} (\hat{c}^\dagger - \hat{c}) 
\label{HamiltonianOptmech}
\end{eqnarray}
in the frame rotating with the laser (pump) frequency $\omega_{\rm L}$, i.e. $\Delta_c=\omega_c-\omega_{\rm L}$. $\omega_c$ is the frequency of the optical cavity mode. Nonclassicalities, e.g entanglement and single-mode nonclassicality~(SMNc) such as squeezing, are determined by noise operators~\cite{simon1994quantum,SimonPRL2000}, i.e. $\delta\hat{c}=\hat{c}-\langle\hat{c}\rangle$ and $\delta\hat{q}=\hat{q}-\langle\hat{q}\rangle$, with $\hat{q}=(\hat{a}^\dagger+\hat{a})/\sqrt{2}$. After the linerization, Langevin equations
\begin{eqnarray}
&& \delta\dot{\hat{q}} = \omega_m \delta\hat{p}
\\
&& \delta\dot{\hat{p}} = -\gamma_m \delta\hat{p} -\omega_m\delta\hat{q} + g(\alpha_c^*\delta\hat{c} + \alpha_c \delta\hat{c}^\dagger) + g_m \hat{\epsilon}_{\rm in}(t) \qquad
\\
&& \delta\dot{\hat{c}} = -(\gamma_c+i\Delta)\delta\hat{c} + ig\alpha_c\delta\hat{q} + g_c \hat{a}_{\rm in}(t)
\end{eqnarray}
become analytically solvable, where $\hat{a}_{\rm in}(t)$ and $\hat{\epsilon}_{\rm in}(t)$ are the optical and mechanical noises, leaking in, from the two vacua~\cite{tasgin2015mutual}. The laser pump is used for increasing the effective coupling between the mirror and cavity mode. $\gamma_{c,m}=\pi D(\omega_{c,m}) g_{c,m}^2$ are the damping rates and $\Delta=\Delta_c-g|q_{s}|^2$ with $q_s$ and $\alpha_c$ are the steady-state values for $\langle\hat{q}\rangle$ and $\langle\hat{c}\rangle$~\cite{agarwal2010electromagnetically,tarhan2013superluminal}.

In this work, in difference to Refs.~\cite{genes&VitaliPRA2008robust,vitaliPRL2007optomechanical}, we are particularly interested in single-mode nonclassicality~(SMNc), e.g. squeezing, of the $\hat{c}$-mode. The reason becomes apparent in the following section. We quantify the SMNc of the cavity mode ---stays Gaussian due to linearization~\cite{genes&VitaliPRA2008robust}--- using a beam-splitter~(BS) approach described in Ref.~\cite{tasgin2015measure} in full details. Shortly, a stronger nonclassical state generates stronger entanglement at a BS output~\cite{ge2015conservation}.

The linearized version of the hamiltonian~(\ref{HamiltonianOptmech}) contains two terms. One of them is the entangler part $\hat{\cal H}_{\rm ent}\propto\hat{c}^\dagger \hat{a}^\dagger + \hat{a}\hat{c}$. The second one is  $\hat{\cal H}_{\rm BS}\propto \hat{a}^\dagger \hat{c}+\hat{c}^\dagger \hat{a}$. The two-mode squeezing hamiltonian~\cite{ScullyZubairyBook} $\hat{\cal H}_{\rm ent}$ generates pure entanglement and  $\hat{\cal H}_{\rm BS}$ interaction just distributes it between different amounts of two-mode entanglement and single-mode nonclassicality, where a conservation holds between the two~\cite{ge2015conservation,arkhipov2016nonclassicality,arkhipov2016interplay,vcernoch2018experimental}. That is,  $\hat{\cal H}_{\rm ent}$ alone, without the presence of $\hat{\cal H}_{\rm BS}$, cannot produce SMNc in the cavity-mode $\hat{c}$.

We tune $\Delta=-\omega_m$. This brings the interaction $\hat{\cal H}_{\rm ent}$ into resonance~\cite{genes&VitaliPRA2008robust,vitaliPRL2007optomechanical} and favors the generation of $\hat{c}$-$\hat{a}$ entanglement. When $\Delta=-\omega_m$, however, instability sets up at a small $g_{\rm st}=\sqrt{\gamma_c\gamma_m/2}$~\cite{genes&VitaliPRA2008robust,vitaliPRL2007optomechanical}. This makes the system unstable before/at the SMNc onsets. For this tuning, action of the  $\hat{\cal H}_{\rm BS}$ is too small due to the off-resonance. We can circumvent this problem by introducing a variable interaction
\begin{equation}
\hat{\cal H}_{\rm int}=\hbar g_1(\hat{c}^\dagger \hat{a}^\dagger + \hat{a}\hat{c}) + \hbar g_2 (\hat{c}^\dagger \hat{a} + \hat{a}^\dagger\hat{c})
\label{Hint_g1g2}
\end{equation}
hamiltonian. Here, we aim to control $\hat{\cal H}_{\rm ent}$ and $\hat{\cal H}_{\rm BS}$ terms independently, via $g_{1,2}$. For $\Delta=-\omega_m$, we increase $g_2$ accordingly \textit{merely} to stabilize the system. That is, smaller values of $g_2$ does not stabilize the system for us to see the increase in the SMNc. 

In Fig.~\ref{fig1}a, we plot the degree of the SMNc of the cavity mode for $\Delta=-\omega_m$ and $g_2=10g_1$. We observe that there is a dramatic increase after $g\geq g_{\rm crt}$. Here $g_{\rm crt}\sim \sqrt{\gamma_c\gamma_m/2}$. In Fig.~\ref{fig1}a, system becomes unstable where the plot ends. We note that $g_2$ is chosen larger only to demonstrate the audience that when the system is stabilized, SMNc and violation of KK relations appear at the same parameters. Although we use $g_2$=10$g_1$ in Fig.~\ref{fig1}, one can appreciate that choice of $\Delta=-\omega_m$, off-resonant, already weakens its act substantially~\cite{genes&VitaliPRA2008robust}.

Next, we leave the second-quantized picture and work with the expectation values of the operators~\cite{agarwal2010electromagnetically,tarhan2013superluminal}, e.g.,
\begin{eqnarray}
c=c_0 + c_+\alpha_p e^{-i\Delta_pt} + c_- \alpha_p^* e^{i\Delta_pt},
\end{eqnarray}
similarly for $q$ and $p$. We assume an extremely small probe field $\alpha_p$, $|\alpha_p|^2$ is the number of probe photons, of frequency $\Delta_p$ in the rotating frame. Since the probe field $\alpha_p$ is extremely small, system can be safely described with the linear response, i.e. $\alpha_p^2\sim 0$. In both treatments second order terms, e.g. $(\delta\hat{q})^2$ and $\alpha_p^2$, are neglected. The nonanalyticities of the transfer function can be obtained from the roots of $c_+=0$~\cite{tasgin2015mutual}, i.e.
\begin{equation}
c_+ \propto [\gamma_c-i(\Delta+\Delta_p)](\Delta_p^2-\omega_m^2+i\gamma_m \Delta_p) - i\omega_m |G|^2 = 0.
\end{equation}
We also checked if the zeros from the denominator of $c_+$ cancels the ones from the nominator and we observed that they are completely different. Denominator does not have a zero in the upper half of the complex frequency place~(CFP). In our cavity system, there appears some nonanalyticities also due to interference~\cite{wangPRA2007theoretical,Zubairy2014counterintuitive,wang2016counterintuitive}, e.g. $-(2\gamma_c \tilde{c}_+ -2)^2  +(2\gamma_c \tilde{c}_+)^2 e^{i2k_pL}+...=0$ with $k_p=\omega_p/c$, which we do not consider here. $c_+=0$ has 3 complex roots $\Delta_p^{(1,2,3)}$, one of them relies in the upper half of the CFP for $g\geq g_{\rm crt}$, see Fig.~\ref{fig1}b.

In Fig.~\ref{fig1}, we clearly observe that for an entangler hamiltonian $\hat{\cal H}_{\rm ent}=\hbar g_1(\hat{c}^\dagger \hat{a}^\dagger + \hat{a}\hat{c})$, violation of KK and SMNc onset at the same critical point. Moreover, this is independent of $\gamma_{c,m}$. In Fig.~\ref{fig1}a, system becomes unstable where the plot ends. We recall that for SMNc of the cavity mode to emerge,  $\hat{\cal H}_{\rm ent}$ interaction is not sufficient. $\hat{\cal H}_{\rm ent}$ only generates the entanglement, but $\hat{\cal H}_{\rm BS}$ distributes the entanglement into  SMNc.

Hence, for the simplest entanglement generator (namely the two-mode squeezing) hamiltonian, SMNc and violation of KK relations appear together~\footnote{Even though probe field and frequency $\Delta_p$ appear explicitly in the calculation of the transfer functions, violation of KK relations is related with the complete frequency response of the transfer functions.}. We note that, the final hamiltonian
\begin{eqnarray}
\hat{\cal H} = \hbar\Delta\hat{c}^\dagger\hat{c} + \hbar\omega_m\hat{a}^\dagger\hat{a} + \hbar g_1(\hat{c}^\dagger \hat{a}^\dagger + \hat{a}\hat{c}) + \hbar g_2 (\hat{c}^\dagger \hat{a} + \hat{a}^\dagger\hat{c})
\label{Hlin_all}
\end{eqnarray}
does {\it not} include any \textit{gain}. We introduce only the losses into Langevin equations. There is no physical phenomenon (restriction) which avoids the achievement of $g_1\geq g_{\rm crt}$ in principle. Our work bases on the solutions of the simplest (standard) exactly solvable nonclassicality~(entanglement and SMNc) generator hamiltonian~(\ref{Hlin_all}). We put the optomechanics hamiltonian as an example physical system where one can approximately obtain the interaction~(\ref{Hint_g1g2}). Other systems could also result similar interaction. For instance, in Ref.~\cite{SERSoptomechanicsNatureNano2016} it is discussed that a radiation pressure like hamiltonian is responsible for the Stokes and anti-Stokes shifts in surface enhanced Raman scattering~(SERS).

Besides its fundamental implications, such a phenomenon can also be used to ``guess" the onset regime for measurements below the standard quantum limit (squeezing) by measuring the transfer functions~\footnote{The credit for this idea belongs to Peter Zoller at IQOQI of Innsbruck.}.


{\small \bf Why SMNc?} Now we are in the right position to state the physical reasoning for considering the degree of SMNc of the cavity field, instead of $\hat{c}-\hat{a}$ entanglement. As explicitly demonstrated in Ref.~\cite{tasgin2017many}, the quasiparticle excitations of an ensemble become nonclassical~(SMNc) when the particles, in the ensemble, are entangled ``collectively". For a better visualization, in Fig.~\ref{fig2} we plot a squeezed ``phonon" wavepacket. The findings of Ref.~\cite{tasgin2017many} state that if the phonon is squeezed~\cite{phonon_squeezing_Science_1997}, vibrational (motional) degree of freedom of atoms are entangled with each other. This happens within the extent of the phonon wavepacket~\cite{tasgin2019Wavepackets}. Hence, the following question would be intriguing. Analogously, can one imagine that, when the light inside the cavity is SMNc~(e.g. squeezed), the different positions in the background spacetime of the cavity (maybe Schwinger pairs at different positions) are entangled~\cite{superfluid_spacetimePRL2014}~\footnote{This can be checked via coupled QED-EFE equations.}? This is the explicit reason we calculate the SMNc of the cavity field instead of the $\hat{c}-\hat{a}$ entanglement. Interestingly, only the SMNc of $\hat{c}$ displays such an increase, not the $\hat{c}$-$\hat{a}$ entanglement. Hence, in this sense, the type of entanglement we treat here can be different than the bipartite entanglement in Refs.~\cite{sonner2013holographic,jensen2013holographic}.

We repeat ourselves: we do not base a theory relying on these coincidences. We only raise them to the attention of the society, which we think they are important issues.

{\bf \small Conclusion}--- The relation between a ``maximally entangled" Schwinger pair and a wormhole connecting the two particles has already been demonstrated explicitly. Here, we question if there could exist a more general connection between entanglement of quantum optics and curvature in general relativity~\cite{sonner2013holographic}. Unlike Refs.~\cite{sonner2013holographic,jensen2013holographic,semenoff2011holographic}, we do not use String Theory, however, nor we can provide a direct derivation like the one stated in Refs.~\cite{sonner2013holographic,jensen2013holographic}.

We consider the possibility of such a relation in a totally different point of view. We provide reasonable ``clues". On one side we show that (a.i) entanglement and (a.ii) violation of Kramers-Kronig relations appear mutually in the simplest exactly solvable hamiltonian. On a second side, Refs.~\cite{hollowood2008refractive,hollowood2007causality,shore2003quantum} show that a small (b.i) curvature in the background spacetime makes the (b.ii) refractive index violate the KK relations in QED.

Keeping in mind that (\texttt{1}) violation of KK relations is not a common (is an unresolved) phenomenon, ({\tt 2}) superluminal propagation is accompanied by violation of KK relations~\cite{tasgin_group_index}, and ({\tt 3}) presence of superluminal propagation and violation of KK relations do not imply the violation of causality in a curved spacetime background~\cite{hollowood2008refractive,hollowood2007causality,shore2003quantum}; makes us consider the possibility of a generalization of the (into weak) entanglement-curvature relation. Actually, this is better than accepting the violation of causality with entanglement in Fig.~\ref{fig1} or in other systems.

We believe that the conjectures we raise and the direction we point out in this work will stimulate new and fundamental works on QED in curved spacetime.




\bibliography{bibliography}

\begin{thebibliography}{64}%
\makeatletter
\providecommand \@ifxundefined [1]{%
 \@ifx{#1\undefined}
}%
\providecommand \@ifnum [1]{%
 \ifnum #1\expandafter \@firstoftwo
 \else \expandafter \@secondoftwo
 \fi
}%
\providecommand \@ifx [1]{%
 \ifx #1\expandafter \@firstoftwo
 \else \expandafter \@secondoftwo
 \fi
}%
\providecommand \natexlab [1]{#1}%
\providecommand \enquote  [1]{``#1''}%
\providecommand \bibnamefont  [1]{#1}%
\providecommand \bibfnamefont [1]{#1}%
\providecommand \citenamefont [1]{#1}%
\providecommand \href@noop [0]{\@secondoftwo}%
\providecommand \href [0]{\begingroup \@sanitize@url \@href}%
\providecommand \@href[1]{\@@startlink{#1}\@@href}%
\providecommand \@@href[1]{\endgroup#1\@@endlink}%
\providecommand \@sanitize@url [0]{\catcode `\\12\catcode `\$12\catcode
  `\&12\catcode `\#12\catcode `\^12\catcode `\_12\catcode `\%12\relax}%
\providecommand \@@startlink[1]{}%
\providecommand \@@endlink[0]{}%
\providecommand \url  [0]{\begingroup\@sanitize@url \@url }%
\providecommand \@url [1]{\endgroup\@href {#1}{\urlprefix }}%
\providecommand \urlprefix  [0]{URL }%
\providecommand \Eprint [0]{\href }%
\providecommand \doibase [0]{http://dx.doi.org/}%
\providecommand \selectlanguage [0]{\@gobble}%
\providecommand \bibinfo  [0]{\@secondoftwo}%
\providecommand \bibfield  [0]{\@secondoftwo}%
\providecommand \translation [1]{[#1]}%
\providecommand \BibitemOpen [0]{}%
\providecommand \bibitemStop [0]{}%
\providecommand \bibitemNoStop [0]{.\EOS\space}%
\providecommand \EOS [0]{\spacefactor3000\relax}%
\providecommand \BibitemShut  [1]{\csname bibitem#1\endcsname}%
\let\auto@bib@innerbib\@empty
\bibitem [{\citenamefont {Eberhard}\ and\ \citenamefont
  {Ross}(1989)}]{controled_SL_communication}%
  \BibitemOpen
  \bibfield  {author} {\bibinfo {author} {\bibfnamefont {Phillippe~H}\
  \bibnamefont {Eberhard}}\ and\ \bibinfo {author} {\bibfnamefont {Ronald~R}\
  \bibnamefont {Ross}},\ }\bibfield  {title} {\enquote {\bibinfo {title}
  {Quantum field theory cannot provide faster-than-light communication},}\
  }\href@noop {} {\bibfield  {journal} {\bibinfo  {journal} {Foundations of
  Physics Letters}\ }\textbf {\bibinfo {volume} {2}},\ \bibinfo {pages}
  {127--149} (\bibinfo {year} {1989})}\BibitemShut {NoStop}%
\bibitem [{\citenamefont {Tasgin}(2017)}]{tasgin2017many}%
  \BibitemOpen
  \bibfield  {author} {\bibinfo {author} {\bibfnamefont {Mehmet~Emre}\
  \bibnamefont {Tasgin}},\ }\bibfield  {title} {\enquote {\bibinfo {title}
  {Many-particle entanglement criterion for superradiantlike states},}\
  }\href@noop {} {\bibfield  {journal} {\bibinfo  {journal} {Physical review
  letters}\ }\textbf {\bibinfo {volume} {119}},\ \bibinfo {pages} {033601}
  (\bibinfo {year} {2017})}\BibitemShut {NoStop}%
\bibitem [{\citenamefont {S{\o}rensen}\ \emph {et~al.}(2001)\citenamefont
  {S{\o}rensen}, \citenamefont {Duan}, \citenamefont {Cirac},\ and\
  \citenamefont {Zoller}}]{sorensen2001Nature}%
  \BibitemOpen
  \bibfield  {author} {\bibinfo {author} {\bibfnamefont {A}~\bibnamefont
  {S{\o}rensen}}, \bibinfo {author} {\bibfnamefont {L-M}\ \bibnamefont {Duan}},
  \bibinfo {author} {\bibfnamefont {JI}~\bibnamefont {Cirac}}, \ and\ \bibinfo
  {author} {\bibfnamefont {Peter}\ \bibnamefont {Zoller}},\ }\bibfield  {title}
  {\enquote {\bibinfo {title} {Many-particle entanglement with bose--einstein
  condensates},}\ }\href@noop {} {\bibfield  {journal} {\bibinfo  {journal}
  {Nature}\ }\textbf {\bibinfo {volume} {409}},\ \bibinfo {pages} {63--66}
  (\bibinfo {year} {2001})}\BibitemShut {NoStop}%
\bibitem [{\citenamefont {Maldacena}\ and\ \citenamefont
  {Susskind}(2013)}]{Susskind2013cool}%
  \BibitemOpen
  \bibfield  {author} {\bibinfo {author} {\bibfnamefont {Juan}\ \bibnamefont
  {Maldacena}}\ and\ \bibinfo {author} {\bibfnamefont {Leonard}\ \bibnamefont
  {Susskind}},\ }\bibfield  {title} {\enquote {\bibinfo {title} {Cool horizons
  for entangled black holes},}\ }\href@noop {} {\bibfield  {journal} {\bibinfo
  {journal} {Fortschritte der Physik}\ }\textbf {\bibinfo {volume} {61}},\
  \bibinfo {pages} {781--811} (\bibinfo {year} {2013})}\BibitemShut {NoStop}%
\bibitem [{\citenamefont {Sonner}(2013)}]{sonner2013holographic}%
  \BibitemOpen
  \bibfield  {author} {\bibinfo {author} {\bibfnamefont {Julian}\ \bibnamefont
  {Sonner}},\ }\bibfield  {title} {\enquote {\bibinfo {title} {Holographic
  schwinger effect and the geometry of entanglement},}\ }\href@noop {}
  {\bibfield  {journal} {\bibinfo  {journal} {Physical Review Letters}\
  }\textbf {\bibinfo {volume} {111}},\ \bibinfo {pages} {211603} (\bibinfo
  {year} {2013})}\BibitemShut {NoStop}%
\bibitem [{\citenamefont {Jensen}\ and\ \citenamefont
  {Karch}(2013)}]{jensen2013holographic}%
  \BibitemOpen
  \bibfield  {author} {\bibinfo {author} {\bibfnamefont {Kristan}\ \bibnamefont
  {Jensen}}\ and\ \bibinfo {author} {\bibfnamefont {Andreas}\ \bibnamefont
  {Karch}},\ }\bibfield  {title} {\enquote {\bibinfo {title} {Holographic dual
  of an einstein-podolsky-rosen pair has a wormhole},}\ }\href@noop {}
  {\bibfield  {journal} {\bibinfo  {journal} {Physical Review Letters}\
  }\textbf {\bibinfo {volume} {111}},\ \bibinfo {pages} {211602} (\bibinfo
  {year} {2013})}\BibitemShut {NoStop}%
\bibitem [{\citenamefont {Maldacena}(1999)}]{maldacena1999large}%
  \BibitemOpen
  \bibfield  {author} {\bibinfo {author} {\bibfnamefont {Juan}\ \bibnamefont
  {Maldacena}},\ }\bibfield  {title} {\enquote {\bibinfo {title} {The large-n
  limit of superconformal field theories and supergravity},}\ }\href@noop {}
  {\bibfield  {journal} {\bibinfo  {journal} {International journal of
  theoretical physics}\ }\textbf {\bibinfo {volume} {38}},\ \bibinfo {pages}
  {1113--1133} (\bibinfo {year} {1999})}\BibitemShut {NoStop}%
\bibitem [{\citenamefont {Witten}(1998)}]{witten1998anti}%
  \BibitemOpen
  \bibfield  {author} {\bibinfo {author} {\bibfnamefont {Edward}\ \bibnamefont
  {Witten}},\ }\bibfield  {title} {\enquote {\bibinfo {title} {Anti de sitter
  space and holography},}\ }\href@noop {} {\bibfield  {journal} {\bibinfo
  {journal} {arXiv preprint hep-th/9802150}\ } (\bibinfo {year}
  {1998})}\BibitemShut {NoStop}%
\bibitem [{\citenamefont {Fujita}\ \emph {et~al.}(2011)\citenamefont {Fujita},
  \citenamefont {Takayanagi},\ and\ \citenamefont {Tonni}}]{fujita2011aspects}%
  \BibitemOpen
  \bibfield  {author} {\bibinfo {author} {\bibfnamefont {Mitsutoshi}\
  \bibnamefont {Fujita}}, \bibinfo {author} {\bibfnamefont {Tadashi}\
  \bibnamefont {Takayanagi}}, \ and\ \bibinfo {author} {\bibfnamefont {Erik}\
  \bibnamefont {Tonni}},\ }\bibfield  {title} {\enquote {\bibinfo {title}
  {Aspects of ads/bcft},}\ }\href@noop {} {\bibfield  {journal} {\bibinfo
  {journal} {Journal of High Energy Physics}\ }\textbf {\bibinfo {volume}
  {2011}},\ \bibinfo {pages} {43} (\bibinfo {year} {2011})}\BibitemShut
  {NoStop}%
\bibitem [{\citenamefont {Xiao}(2008)}]{xiao2008exact}%
  \BibitemOpen
  \bibfield  {author} {\bibinfo {author} {\bibfnamefont {Bo-Wen}\ \bibnamefont
  {Xiao}},\ }\bibfield  {title} {\enquote {\bibinfo {title} {On the exact
  solution of the accelerating string in ads5 space},}\ }\href@noop {}
  {\bibfield  {journal} {\bibinfo  {journal} {Physics Letters B}\ }\textbf
  {\bibinfo {volume} {665}},\ \bibinfo {pages} {173--177} (\bibinfo {year}
  {2008})}\BibitemShut {NoStop}%
\bibitem [{\citenamefont {Semenoff}\ and\ \citenamefont
  {Zarembo}(2011)}]{semenoff2011holographic}%
  \BibitemOpen
  \bibfield  {author} {\bibinfo {author} {\bibfnamefont {Gordon~W}\
  \bibnamefont {Semenoff}}\ and\ \bibinfo {author} {\bibfnamefont {Konstantin}\
  \bibnamefont {Zarembo}},\ }\bibfield  {title} {\enquote {\bibinfo {title}
  {Holographic schwinger effect},}\ }\href@noop {} {\bibfield  {journal}
  {\bibinfo  {journal} {Physical review letters}\ }\textbf {\bibinfo {volume}
  {107}},\ \bibinfo {pages} {171601} (\bibinfo {year} {2011})}\BibitemShut
  {NoStop}%
\bibitem [{\citenamefont {Dunne}\ and\ \citenamefont
  {Schubert}(2005)}]{dunne2005worldline}%
  \BibitemOpen
  \bibfield  {author} {\bibinfo {author} {\bibfnamefont {Gerald~V}\
  \bibnamefont {Dunne}}\ and\ \bibinfo {author} {\bibfnamefont {Christian}\
  \bibnamefont {Schubert}},\ }\bibfield  {title} {\enquote {\bibinfo {title}
  {Worldline instantons and pair production in inhomogenous fields},}\
  }\href@noop {} {\bibfield  {journal} {\bibinfo  {journal} {Physical Review
  D}\ }\textbf {\bibinfo {volume} {72}},\ \bibinfo {pages} {105004} (\bibinfo
  {year} {2005})}\BibitemShut {NoStop}%
\bibitem [{\citenamefont {Dunne}\ \emph {et~al.}(2006)\citenamefont {Dunne},
  \citenamefont {Wang}, \citenamefont {Gies},\ and\ \citenamefont
  {Schubert}}]{dunne2006worldline}%
  \BibitemOpen
  \bibfield  {author} {\bibinfo {author} {\bibfnamefont {Gerald~V}\
  \bibnamefont {Dunne}}, \bibinfo {author} {\bibfnamefont {Qing-hai}\
  \bibnamefont {Wang}}, \bibinfo {author} {\bibfnamefont {Holger}\ \bibnamefont
  {Gies}}, \ and\ \bibinfo {author} {\bibfnamefont {Christian}\ \bibnamefont
  {Schubert}},\ }\bibfield  {title} {\enquote {\bibinfo {title} {Worldline
  instantons and the fluctuation prefactor},}\ }\href@noop {} {\bibfield
  {journal} {\bibinfo  {journal} {Physical Review D}\ }\textbf {\bibinfo
  {volume} {73}},\ \bibinfo {pages} {065028} (\bibinfo {year}
  {2006})}\BibitemShut {NoStop}%
\bibitem [{\citenamefont {Kim}\ \emph {et~al.}(1997)\citenamefont {Kim},
  \citenamefont {Lee},\ and\ \citenamefont {Myung}}]{kim1997classical}%
  \BibitemOpen
  \bibfield  {author} {\bibinfo {author} {\bibfnamefont {Jin~Young}\
  \bibnamefont {Kim}}, \bibinfo {author} {\bibfnamefont {HW}~\bibnamefont
  {Lee}}, \ and\ \bibinfo {author} {\bibfnamefont {YS}~\bibnamefont {Myung}},\
  }\bibfield  {title} {\enquote {\bibinfo {title} {Classical instanton and
  wormhole solutions of type iib string theory},}\ }\href@noop {} {\bibfield
  {journal} {\bibinfo  {journal} {Physics Letters B}\ }\textbf {\bibinfo
  {volume} {400}},\ \bibinfo {pages} {32--36} (\bibinfo {year}
  {1997})}\BibitemShut {NoStop}%
\bibitem [{\citenamefont {Gutperle}\ and\ \citenamefont
  {Sabra}(2002)}]{gutperle2002instantons}%
  \BibitemOpen
  \bibfield  {author} {\bibinfo {author} {\bibfnamefont {Michael}\ \bibnamefont
  {Gutperle}}\ and\ \bibinfo {author} {\bibfnamefont {Wafic}\ \bibnamefont
  {Sabra}},\ }\bibfield  {title} {\enquote {\bibinfo {title} {Instantons and
  wormholes in minkowski and (a) ds spaces},}\ }\href@noop {} {\bibfield
  {journal} {\bibinfo  {journal} {Nuclear Physics B}\ }\textbf {\bibinfo
  {volume} {647}},\ \bibinfo {pages} {344--356} (\bibinfo {year}
  {2002})}\BibitemShut {NoStop}%
\bibitem [{\citenamefont {Gorsky}\ \emph {et~al.}(2002)\citenamefont {Gorsky},
  \citenamefont {Saraikin},\ and\ \citenamefont
  {Selivanov}}]{gorsky2002schwinger}%
  \BibitemOpen
  \bibfield  {author} {\bibinfo {author} {\bibfnamefont {AS}~\bibnamefont
  {Gorsky}}, \bibinfo {author} {\bibfnamefont {KA}~\bibnamefont {Saraikin}}, \
  and\ \bibinfo {author} {\bibfnamefont {KG}~\bibnamefont {Selivanov}},\
  }\bibfield  {title} {\enquote {\bibinfo {title} {Schwinger type processes via
  branes and their gravity duals},}\ }\href@noop {} {\bibfield  {journal}
  {\bibinfo  {journal} {Nuclear Physics B}\ }\textbf {\bibinfo {volume}
  {628}},\ \bibinfo {pages} {270--294} (\bibinfo {year} {2002})}\BibitemShut
  {NoStop}%
\bibitem [{\citenamefont {Hollowood}\ and\ \citenamefont
  {Shore}(2008)}]{hollowood2008refractive}%
  \BibitemOpen
  \bibfield  {author} {\bibinfo {author} {\bibfnamefont {Timothy~J}\
  \bibnamefont {Hollowood}}\ and\ \bibinfo {author} {\bibfnamefont {Graham~M}\
  \bibnamefont {Shore}},\ }\bibfield  {title} {\enquote {\bibinfo {title} {The
  refractive index of curved spacetime: the fate of causality in qed},}\
  }\href@noop {} {\bibfield  {journal} {\bibinfo  {journal} {Nuclear physics
  B}\ }\textbf {\bibinfo {volume} {795}},\ \bibinfo {pages} {138--171}
  (\bibinfo {year} {2008})}\BibitemShut {NoStop}%
\bibitem [{\citenamefont {Hollowood}\ and\ \citenamefont
  {Shore}(2007)}]{hollowood2007causality}%
  \BibitemOpen
  \bibfield  {author} {\bibinfo {author} {\bibfnamefont {Timothy~J}\
  \bibnamefont {Hollowood}}\ and\ \bibinfo {author} {\bibfnamefont {Graham~M}\
  \bibnamefont {Shore}},\ }\bibfield  {title} {\enquote {\bibinfo {title}
  {Causality and micro-causality in curved spacetime},}\ }\href@noop {}
  {\bibfield  {journal} {\bibinfo  {journal} {Physics Letters B}\ }\textbf
  {\bibinfo {volume} {655}},\ \bibinfo {pages} {67--74} (\bibinfo {year}
  {2007})}\BibitemShut {NoStop}%
\bibitem [{\citenamefont {Josse}\ \emph {et~al.}(2004)\citenamefont {Josse},
  \citenamefont {Dantan}, \citenamefont {Bramati},\ and\ \citenamefont
  {Giacobino}}]{josse2004entanglement}%
  \BibitemOpen
  \bibfield  {author} {\bibinfo {author} {\bibfnamefont {Vincent}\ \bibnamefont
  {Josse}}, \bibinfo {author} {\bibfnamefont {Aur{\'e}lien}\ \bibnamefont
  {Dantan}}, \bibinfo {author} {\bibfnamefont {Alberto}\ \bibnamefont
  {Bramati}}, \ and\ \bibinfo {author} {\bibfnamefont {Elisabeth}\ \bibnamefont
  {Giacobino}},\ }\bibfield  {title} {\enquote {\bibinfo {title} {Entanglement
  and squeezing in a two-mode system: theory and experiment},}\ }\href@noop {}
  {\bibfield  {journal} {\bibinfo  {journal} {Journal of Optics B: Quantum and
  Semiclassical Optics}\ }\textbf {\bibinfo {volume} {6}},\ \bibinfo {pages}
  {S532} (\bibinfo {year} {2004})}\BibitemShut {NoStop}%
\bibitem [{\citenamefont {Liberati}\ and\ \citenamefont
  {Maccione}(2014)}]{superfluid_spacetimePRL2014}%
  \BibitemOpen
  \bibfield  {author} {\bibinfo {author} {\bibfnamefont {Stefano}\ \bibnamefont
  {Liberati}}\ and\ \bibinfo {author} {\bibfnamefont {Luca}\ \bibnamefont
  {Maccione}},\ }\bibfield  {title} {\enquote {\bibinfo {title} {Astrophysical
  constraints on planck scale dissipative phenomena},}\ }\href@noop {}
  {\bibfield  {journal} {\bibinfo  {journal} {Physical Review Letters}\
  }\textbf {\bibinfo {volume} {112}},\ \bibinfo {pages} {151301} (\bibinfo
  {year} {2014})}\BibitemShut {NoStop}%
\bibitem [{\citenamefont {Feynman}(1950)}]{feynman1950mathematical}%
  \BibitemOpen
  \bibfield  {author} {\bibinfo {author} {\bibfnamefont {Richard~Phillips}\
  \bibnamefont {Feynman}},\ }\bibfield  {title} {\enquote {\bibinfo {title}
  {Mathematical formulation of the quantum theory of electromagnetic
  interaction},}\ }\href@noop {} {\bibfield  {journal} {\bibinfo  {journal}
  {Physical Review}\ }\textbf {\bibinfo {volume} {80}},\ \bibinfo {pages} {440}
  (\bibinfo {year} {1950})}\BibitemShut {NoStop}%
\bibitem [{\citenamefont {Schwinger}(1951)}]{schwinger1951theory}%
  \BibitemOpen
  \bibfield  {author} {\bibinfo {author} {\bibfnamefont {Julian}\ \bibnamefont
  {Schwinger}},\ }\bibfield  {title} {\enquote {\bibinfo {title} {The theory of
  quantized fields. i},}\ }\href@noop {} {\bibfield  {journal} {\bibinfo
  {journal} {Physical Review}\ }\textbf {\bibinfo {volume} {82}},\ \bibinfo
  {pages} {914} (\bibinfo {year} {1951})}\BibitemShut {NoStop}%
\bibitem [{\citenamefont {Bastianelli}\ and\ \citenamefont
  {Zirotti}(2002)}]{bastianelli2002worldline}%
  \BibitemOpen
  \bibfield  {author} {\bibinfo {author} {\bibfnamefont {Fiorenzo}\
  \bibnamefont {Bastianelli}}\ and\ \bibinfo {author} {\bibfnamefont {Andrea}\
  \bibnamefont {Zirotti}},\ }\bibfield  {title} {\enquote {\bibinfo {title}
  {Worldline formalism in a gravitational background},}\ }\href@noop {}
  {\bibfield  {journal} {\bibinfo  {journal} {Nuclear Physics B}\ }\textbf
  {\bibinfo {volume} {642}},\ \bibinfo {pages} {372--388} (\bibinfo {year}
  {2002})}\BibitemShut {NoStop}%
\bibitem [{\citenamefont {Schubert}(2001)}]{schubert2001perturbative}%
  \BibitemOpen
  \bibfield  {author} {\bibinfo {author} {\bibfnamefont {Christian}\
  \bibnamefont {Schubert}},\ }\bibfield  {title} {\enquote {\bibinfo {title}
  {Perturbative quantum field theory in the string-inspired formalism},}\
  }\href@noop {} {\bibfield  {journal} {\bibinfo  {journal} {Physics Reports}\
  }\textbf {\bibinfo {volume} {355}},\ \bibinfo {pages} {73--234} (\bibinfo
  {year} {2001})}\BibitemShut {NoStop}%
\bibitem [{\citenamefont {Shore}(2003)}]{shore2003quantum}%
  \BibitemOpen
  \bibfield  {author} {\bibinfo {author} {\bibfnamefont {Graham~M}\
  \bibnamefont {Shore}},\ }\bibfield  {title} {\enquote {\bibinfo {title}
  {Quantum gravitational optics},}\ }\href@noop {} {\bibfield  {journal}
  {\bibinfo  {journal} {Contemporary Physics}\ }\textbf {\bibinfo {volume}
  {44}},\ \bibinfo {pages} {503--521} (\bibinfo {year} {2003})}\BibitemShut
  {NoStop}%
\bibitem [{\citenamefont {Drummond}\ and\ \citenamefont
  {Hathrell}(1980)}]{drummond1980qed}%
  \BibitemOpen
  \bibfield  {author} {\bibinfo {author} {\bibfnamefont {Ian~T}\ \bibnamefont
  {Drummond}}\ and\ \bibinfo {author} {\bibfnamefont {SJ}~\bibnamefont
  {Hathrell}},\ }\bibfield  {title} {\enquote {\bibinfo {title} {Qed vacuum
  polarization in a background gravitational field and its effect on the
  velocity of photons},}\ }\href@noop {} {\bibfield  {journal} {\bibinfo
  {journal} {Physical Review D}\ }\textbf {\bibinfo {volume} {22}},\ \bibinfo
  {pages} {343} (\bibinfo {year} {1980})}\BibitemShut {NoStop}%
\bibitem [{\citenamefont {Bertotti}\ and\ \citenamefont
  {Grishchuk}(1990)}]{bertotti1990strong}%
  \BibitemOpen
  \bibfield  {author} {\bibinfo {author} {\bibfnamefont {B}~\bibnamefont
  {Bertotti}}\ and\ \bibinfo {author} {\bibfnamefont {LP}~\bibnamefont
  {Grishchuk}},\ }\bibfield  {title} {\enquote {\bibinfo {title} {The strong
  equivalence principle},}\ }\href@noop {} {\bibfield  {journal} {\bibinfo
  {journal} {Classical and Quantum Gravity}\ }\textbf {\bibinfo {volume} {7}},\
  \bibinfo {pages} {1733} (\bibinfo {year} {1990})}\BibitemShut {NoStop}%
\bibitem [{\citenamefont {Shore}(2002)}]{shore2002faster}%
  \BibitemOpen
  \bibfield  {author} {\bibinfo {author} {\bibfnamefont {Graham~M}\
  \bibnamefont {Shore}},\ }\bibfield  {title} {\enquote {\bibinfo {title}
  {Faster than light photons in gravitational fields ii.: Dispersion and vacuum
  polarisation},}\ }\href@noop {} {\bibfield  {journal} {\bibinfo  {journal}
  {Nuclear Physics B}\ }\textbf {\bibinfo {volume} {633}},\ \bibinfo {pages}
  {271--294} (\bibinfo {year} {2002})}\BibitemShut {NoStop}%
\bibitem [{\citenamefont {Bruschi}(2016)}]{bruschi2016weight}%
  \BibitemOpen
  \bibfield  {author} {\bibinfo {author} {\bibfnamefont {David~Edward}\
  \bibnamefont {Bruschi}},\ }\bibfield  {title} {\enquote {\bibinfo {title} {On
  the weight of entanglement},}\ }\href@noop {} {\bibfield  {journal} {\bibinfo
   {journal} {Physics Letters B}\ }\textbf {\bibinfo {volume} {754}},\ \bibinfo
  {pages} {182--186} (\bibinfo {year} {2016})}\BibitemShut {NoStop}%
\bibitem [{\citenamefont {Tarhan}\ and\ \citenamefont
  {Tasgin}(2015)}]{tasgin2015mutual}%
  \BibitemOpen
  \bibfield  {author} {\bibinfo {author} {\bibfnamefont {Devrim}\ \bibnamefont
  {Tarhan}}\ and\ \bibinfo {author} {\bibfnamefont {Mehmet~Emre}\ \bibnamefont
  {Tasgin}},\ }\bibfield  {title} {\enquote {\bibinfo {title} {Mutual emergence
  of noncausal optical response and nonclassicality in an optomechanical
  system},}\ }\href@noop {} {\bibfield  {journal} {\bibinfo  {journal} {arXiv
  preprint arXiv:1502.01294}\ } (\bibinfo {year} {2015})}\BibitemShut {NoStop}%
\bibitem [{\citenamefont {Beck}\ \emph {et~al.}(1991)\citenamefont {Beck},
  \citenamefont {Walmsley},\ and\ \citenamefont {Kafka}}]{beck1991group}%
  \BibitemOpen
  \bibfield  {author} {\bibinfo {author} {\bibfnamefont {M}~\bibnamefont
  {Beck}}, \bibinfo {author} {\bibfnamefont {IA}~\bibnamefont {Walmsley}}, \
  and\ \bibinfo {author} {\bibfnamefont {JD}~\bibnamefont {Kafka}},\ }\bibfield
   {title} {\enquote {\bibinfo {title} {Group delay measurements of optical
  components near 800 nm},}\ }\href@noop {} {\bibfield  {journal} {\bibinfo
  {journal} {IEEE journal of quantum electronics}\ }\textbf {\bibinfo {volume}
  {27}},\ \bibinfo {pages} {2074--2081} (\bibinfo {year} {1991})}\BibitemShut
  {NoStop}%
\bibitem [{\citenamefont {Wang}(2002)}]{wang2002causal}%
  \BibitemOpen
  \bibfield  {author} {\bibinfo {author} {\bibfnamefont {LJ}~\bibnamefont
  {Wang}},\ }\bibfield  {title} {\enquote {\bibinfo {title} {Causal
  “all-pass” filters and kramers--kronig relations},}\ }\href@noop {}
  {\bibfield  {journal} {\bibinfo  {journal} {Optics communications}\ }\textbf
  {\bibinfo {volume} {213}},\ \bibinfo {pages} {27--32} (\bibinfo {year}
  {2002})}\BibitemShut {NoStop}%
\bibitem [{\citenamefont {Stern}\ and\ \citenamefont
  {Levy}(2012)}]{stern2012transmission}%
  \BibitemOpen
  \bibfield  {author} {\bibinfo {author} {\bibfnamefont {Liron}\ \bibnamefont
  {Stern}}\ and\ \bibinfo {author} {\bibfnamefont {Uriel}\ \bibnamefont
  {Levy}},\ }\bibfield  {title} {\enquote {\bibinfo {title} {Transmission and
  time delay properties of an integrated system consisting of atomic vapor
  cladding on top of a micro ring resonator},}\ }\href@noop {} {\bibfield
  {journal} {\bibinfo  {journal} {Optics express}\ }\textbf {\bibinfo {volume}
  {20}},\ \bibinfo {pages} {28082--28093} (\bibinfo {year} {2012})}\BibitemShut
  {NoStop}%
\bibitem [{\citenamefont {Chiao}(1999)}]{chiao1999tunneling}%
  \BibitemOpen
  \bibfield  {author} {\bibinfo {author} {\bibfnamefont {Raymond~Y}\
  \bibnamefont {Chiao}},\ }\bibfield  {title} {\enquote {\bibinfo {title}
  {Tunneling times and superluminality: A tutorial},}\ }in\ \href@noop {}
  {\emph {\bibinfo {booktitle} {AIP Conference Proceedings}}},\ Vol.\ \bibinfo
  {volume} {461}\ (\bibinfo {organization} {AIP},\ \bibinfo {year} {1999})\
  pp.\ \bibinfo {pages} {3--13}\BibitemShut {NoStop}%
\bibitem [{\citenamefont {Davies}(2005)}]{davies2005quantum}%
  \BibitemOpen
  \bibfield  {author} {\bibinfo {author} {\bibfnamefont {Paul Charles~William}\
  \bibnamefont {Davies}},\ }\bibfield  {title} {\enquote {\bibinfo {title}
  {Quantum tunneling time},}\ }\href@noop {} {\bibfield  {journal} {\bibinfo
  {journal} {American journal of physics}\ }\textbf {\bibinfo {volume} {73}},\
  \bibinfo {pages} {23--27} (\bibinfo {year} {2005})}\BibitemShut {NoStop}%
\bibitem [{\citenamefont {Wang}\ and\ \citenamefont
  {Xiong}(2007)}]{wangPRA2007theoretical}%
  \BibitemOpen
  \bibfield  {author} {\bibinfo {author} {\bibfnamefont {Zhi-Yong}\
  \bibnamefont {Wang}}\ and\ \bibinfo {author} {\bibfnamefont {Cai-Dong}\
  \bibnamefont {Xiong}},\ }\bibfield  {title} {\enquote {\bibinfo {title}
  {Theoretical evidence for the superluminality of evanescent modes},}\
  }\href@noop {} {\bibfield  {journal} {\bibinfo  {journal} {Physical Review
  A}\ }\textbf {\bibinfo {volume} {75}},\ \bibinfo {pages} {042105} (\bibinfo
  {year} {2007})}\BibitemShut {NoStop}%
\bibitem [{\citenamefont {Winful}(2003{\natexlab{a}})}]{winfulNature2003}%
  \BibitemOpen
  \bibfield  {author} {\bibinfo {author} {\bibfnamefont {Herbert~G}\
  \bibnamefont {Winful}},\ }\bibfield  {title} {\enquote {\bibinfo {title}
  {Optics (communication arising): Mechanism for'superluminal'tunnelling},}\
  }\href@noop {} {\bibfield  {journal} {\bibinfo  {journal} {Nature}\ }\textbf
  {\bibinfo {volume} {424}},\ \bibinfo {pages} {638} (\bibinfo {year}
  {2003}{\natexlab{a}})}\BibitemShut {NoStop}%
\bibitem [{\citenamefont {Winful}(2003{\natexlab{b}})}]{winfulPRL2003}%
  \BibitemOpen
  \bibfield  {author} {\bibinfo {author} {\bibfnamefont {Herbert~G}\
  \bibnamefont {Winful}},\ }\bibfield  {title} {\enquote {\bibinfo {title}
  {Nature of “superluminal" barrier tunneling},}\ }\href@noop {} {\bibfield
  {journal} {\bibinfo  {journal} {Physical review letters}\ }\textbf {\bibinfo
  {volume} {90}},\ \bibinfo {pages} {023901} (\bibinfo {year}
  {2003}{\natexlab{b}})}\BibitemShut {NoStop}%
\bibitem [{\citenamefont {Hegerfeldt}(1998)}]{hegerfeldt1998instantaneous}%
  \BibitemOpen
  \bibfield  {author} {\bibinfo {author} {\bibfnamefont {Gerhard~C}\
  \bibnamefont {Hegerfeldt}},\ }\bibfield  {title} {\enquote {\bibinfo {title}
  {Instantaneous spreading and einstein causality in quantum theory},}\
  }\href@noop {} {\bibfield  {journal} {\bibinfo  {journal} {Annalen der
  Physik}\ }\textbf {\bibinfo {volume} {7}},\ \bibinfo {pages} {716--725}
  (\bibinfo {year} {1998})}\BibitemShut {NoStop}%
\bibitem [{\citenamefont {Hegerfeldt}(1974)}]{hegerfeldtPRD1974remark}%
  \BibitemOpen
  \bibfield  {author} {\bibinfo {author} {\bibfnamefont {Gerhard~C}\
  \bibnamefont {Hegerfeldt}},\ }\bibfield  {title} {\enquote {\bibinfo {title}
  {Remark on causality and particle localization},}\ }\href@noop {} {\bibfield
  {journal} {\bibinfo  {journal} {Physical Review D}\ }\textbf {\bibinfo
  {volume} {10}},\ \bibinfo {pages} {3320} (\bibinfo {year}
  {1974})}\BibitemShut {NoStop}%
\bibitem [{\citenamefont {Perez}\ and\ \citenamefont
  {Wilde}(1977)}]{perezPRD1977localization}%
  \BibitemOpen
  \bibfield  {author} {\bibinfo {author} {\bibfnamefont {J~Fernando}\
  \bibnamefont {Perez}}\ and\ \bibinfo {author} {\bibfnamefont {Ivan~F}\
  \bibnamefont {Wilde}},\ }\bibfield  {title} {\enquote {\bibinfo {title}
  {Localization and causality in relativistic quantum mechanics},}\ }\href@noop
  {} {\bibfield  {journal} {\bibinfo  {journal} {Physical Review D}\ }\textbf
  {\bibinfo {volume} {16}},\ \bibinfo {pages} {315} (\bibinfo {year}
  {1977})}\BibitemShut {NoStop}%
\bibitem [{\citenamefont {Hegerfeldt}\ and\ \citenamefont
  {Ruijsenaars}(1980)}]{hegerfeldtPRD1980remarks}%
  \BibitemOpen
  \bibfield  {author} {\bibinfo {author} {\bibfnamefont {Gerhard~C}\
  \bibnamefont {Hegerfeldt}}\ and\ \bibinfo {author} {\bibfnamefont {Simon~NM}\
  \bibnamefont {Ruijsenaars}},\ }\bibfield  {title} {\enquote {\bibinfo {title}
  {Remarks on causality, localization, and spreading of wave packets},}\
  }\href@noop {} {\bibfield  {journal} {\bibinfo  {journal} {Physical Review
  D}\ }\textbf {\bibinfo {volume} {22}},\ \bibinfo {pages} {377} (\bibinfo
  {year} {1980})}\BibitemShut {NoStop}%
\bibitem [{\citenamefont {{M. E. Tasgin}}(2019)}]{tasgin_group_index}%
  \BibitemOpen
  \bibfield  {author} {\bibinfo {author} {\bibnamefont {{M. E. Tasgin}}},\
  }\href@noop {} {\enquote {\bibinfo {title} {{A Lorentzian group-index
  violates Kramers-Kronig relations}},}\ } (\bibinfo {year} {2019}),\ \bibinfo
  {note} {see on Researchgate or arXiv with the title}\BibitemShut {NoStop}%
\bibitem [{\citenamefont {Chu}\ and\ \citenamefont
  {Wong}(1982)}]{ChuPRL1982SL}%
  \BibitemOpen
  \bibfield  {author} {\bibinfo {author} {\bibfnamefont {S}~\bibnamefont
  {Chu}}\ and\ \bibinfo {author} {\bibfnamefont {S}~\bibnamefont {Wong}},\
  }\bibfield  {title} {\enquote {\bibinfo {title} {Linear pulse propagation in
  an absorbing medium},}\ }\href@noop {} {\bibfield  {journal} {\bibinfo
  {journal} {Physical Review Letters}\ }\textbf {\bibinfo {volume} {48}},\
  \bibinfo {pages} {738} (\bibinfo {year} {1982})}\BibitemShut {NoStop}%
\bibitem [{\citenamefont {Jackson}(1999)}]{Jackson_book}%
  \BibitemOpen
  \bibfield  {author} {\bibinfo {author} {\bibfnamefont {John~David}\
  \bibnamefont {Jackson}},\ }\href {http://cdsweb.cern.ch/record/490457} {\emph
  {\bibinfo {title} {Classical electrodynamics}}},\ \bibinfo {edition} {3rd}\
  ed.\ (\bibinfo  {publisher} {Wiley},\ \bibinfo {address} {New York, {NY}},\
  \bibinfo {year} {1999})\BibitemShut {NoStop}%
\bibitem [{\citenamefont {Wang}\ \emph {et~al.}(2014)\citenamefont {Wang},
  \citenamefont {Wang}, \citenamefont {Al-Amri}, \citenamefont {Zhu},\ and\
  \citenamefont {Zubairy}}]{Zubairy2014counterintuitive}%
  \BibitemOpen
  \bibfield  {author} {\bibinfo {author} {\bibfnamefont {Li-Gang}\ \bibnamefont
  {Wang}}, \bibinfo {author} {\bibfnamefont {Lin}\ \bibnamefont {Wang}},
  \bibinfo {author} {\bibfnamefont {M}~\bibnamefont {Al-Amri}}, \bibinfo
  {author} {\bibfnamefont {Shi-Yao}\ \bibnamefont {Zhu}}, \ and\ \bibinfo
  {author} {\bibfnamefont {M~Suhail}\ \bibnamefont {Zubairy}},\ }\bibfield
  {title} {\enquote {\bibinfo {title} {Counterintuitive dispersion violating
  kramers-kronig relations in gain slabs},}\ }\href@noop {} {\bibfield
  {journal} {\bibinfo  {journal} {Physical review letters}\ }\textbf {\bibinfo
  {volume} {112}},\ \bibinfo {pages} {233601} (\bibinfo {year}
  {2014})}\BibitemShut {NoStop}%
\bibitem [{\citenamefont {Wang}\ \emph {et~al.}(2016)\citenamefont {Wang},
  \citenamefont {Wang}, \citenamefont {Ye}, \citenamefont {Al-Amri},
  \citenamefont {Zhu},\ and\ \citenamefont
  {Zubairy}}]{wang2016counterintuitive}%
  \BibitemOpen
  \bibfield  {author} {\bibinfo {author} {\bibfnamefont {Lin}\ \bibnamefont
  {Wang}}, \bibinfo {author} {\bibfnamefont {Li-Gang}\ \bibnamefont {Wang}},
  \bibinfo {author} {\bibfnamefont {Lin-Hua}\ \bibnamefont {Ye}}, \bibinfo
  {author} {\bibfnamefont {M}~\bibnamefont {Al-Amri}}, \bibinfo {author}
  {\bibfnamefont {Shi-Yao}\ \bibnamefont {Zhu}}, \ and\ \bibinfo {author}
  {\bibfnamefont {M~Suhail}\ \bibnamefont {Zubairy}},\ }\bibfield  {title}
  {\enquote {\bibinfo {title} {Counterintuitive dispersion effect near surface
  plasmon resonances in otto structures},}\ }\href@noop {} {\bibfield
  {journal} {\bibinfo  {journal} {Physical Review A}\ }\textbf {\bibinfo
  {volume} {94}},\ \bibinfo {pages} {013806} (\bibinfo {year}
  {2016})}\BibitemShut {NoStop}%
\bibitem [{\citenamefont {Wang}\ and\ \citenamefont
  {Zhu}(2006)}]{wang2006superluminal}%
  \BibitemOpen
  \bibfield  {author} {\bibinfo {author} {\bibfnamefont {Li-Gang}\ \bibnamefont
  {Wang}}\ and\ \bibinfo {author} {\bibfnamefont {Shi-Yao}\ \bibnamefont
  {Zhu}},\ }\bibfield  {title} {\enquote {\bibinfo {title} {Superluminal pulse
  reflection from a weakly absorbing dielectric slab},}\ }\href@noop {}
  {\bibfield  {journal} {\bibinfo  {journal} {Optics letters}\ }\textbf
  {\bibinfo {volume} {31}},\ \bibinfo {pages} {2223--2225} (\bibinfo {year}
  {2006})}\BibitemShut {NoStop}%
\bibitem [{\citenamefont {Genes}\ \emph {et~al.}(2008)\citenamefont {Genes},
  \citenamefont {Mari}, \citenamefont {Tombesi},\ and\ \citenamefont
  {Vitali}}]{genes&VitaliPRA2008robust}%
  \BibitemOpen
  \bibfield  {author} {\bibinfo {author} {\bibfnamefont {C}~\bibnamefont
  {Genes}}, \bibinfo {author} {\bibfnamefont {A}~\bibnamefont {Mari}}, \bibinfo
  {author} {\bibfnamefont {P}~\bibnamefont {Tombesi}}, \ and\ \bibinfo {author}
  {\bibfnamefont {D}~\bibnamefont {Vitali}},\ }\bibfield  {title} {\enquote
  {\bibinfo {title} {Robust entanglement of a micromechanical resonator with
  output optical fields},}\ }\href@noop {} {\bibfield  {journal} {\bibinfo
  {journal} {Physical Review A}\ }\textbf {\bibinfo {volume} {78}},\ \bibinfo
  {pages} {032316} (\bibinfo {year} {2008})}\BibitemShut {NoStop}%
\bibitem [{\citenamefont {Vitali}\ \emph {et~al.}(2007)\citenamefont {Vitali},
  \citenamefont {Gigan}, \citenamefont {Ferreira}, \citenamefont {B{\"o}hm},
  \citenamefont {Tombesi}, \citenamefont {Guerreiro}, \citenamefont {Vedral},
  \citenamefont {Zeilinger},\ and\ \citenamefont
  {Aspelmeyer}}]{vitaliPRL2007optomechanical}%
  \BibitemOpen
  \bibfield  {author} {\bibinfo {author} {\bibfnamefont {David}\ \bibnamefont
  {Vitali}}, \bibinfo {author} {\bibfnamefont {Sylvain}\ \bibnamefont {Gigan}},
  \bibinfo {author} {\bibfnamefont {Anderson}\ \bibnamefont {Ferreira}},
  \bibinfo {author} {\bibfnamefont {HR}~\bibnamefont {B{\"o}hm}}, \bibinfo
  {author} {\bibfnamefont {Paolo}\ \bibnamefont {Tombesi}}, \bibinfo {author}
  {\bibfnamefont {Ariel}\ \bibnamefont {Guerreiro}}, \bibinfo {author}
  {\bibfnamefont {Vlatko}\ \bibnamefont {Vedral}}, \bibinfo {author}
  {\bibfnamefont {Anton}\ \bibnamefont {Zeilinger}}, \ and\ \bibinfo {author}
  {\bibfnamefont {Markus}\ \bibnamefont {Aspelmeyer}},\ }\bibfield  {title}
  {\enquote {\bibinfo {title} {Optomechanical entanglement between a movable
  mirror and a cavity field},}\ }\href@noop {} {\bibfield  {journal} {\bibinfo
  {journal} {Physical Review Letters}\ }\textbf {\bibinfo {volume} {98}},\
  \bibinfo {pages} {030405} (\bibinfo {year} {2007})}\BibitemShut {NoStop}%
\bibitem [{\citenamefont {Marquardt}\ and\ \citenamefont
  {Girvin}(2009)}]{marquardt2009optomechanics}%
  \BibitemOpen
  \bibfield  {author} {\bibinfo {author} {\bibfnamefont {Florian}\ \bibnamefont
  {Marquardt}}\ and\ \bibinfo {author} {\bibfnamefont {Steven~M}\ \bibnamefont
  {Girvin}},\ }\bibfield  {title} {\enquote {\bibinfo {title} {Optomechanics (a
  brief review)},}\ }\href@noop {} {\bibfield  {journal} {\bibinfo  {journal}
  {arXiv preprint arXiv:0905.0566}\ } (\bibinfo {year} {2009})}\BibitemShut
  {NoStop}%
\bibitem [{\citenamefont {Simon}\ \emph {et~al.}(1994)\citenamefont {Simon},
  \citenamefont {Mukunda},\ and\ \citenamefont {Dutta}}]{simon1994quantum}%
  \BibitemOpen
  \bibfield  {author} {\bibinfo {author} {\bibfnamefont {R}~\bibnamefont
  {Simon}}, \bibinfo {author} {\bibfnamefont {N}~\bibnamefont {Mukunda}}, \
  and\ \bibinfo {author} {\bibfnamefont {Biswadeb}\ \bibnamefont {Dutta}},\
  }\bibfield  {title} {\enquote {\bibinfo {title} {Quantum-noise matrix for
  multimode systems: U (n) invariance, squeezing, and normal forms},}\
  }\href@noop {} {\bibfield  {journal} {\bibinfo  {journal} {Physical Review
  A}\ }\textbf {\bibinfo {volume} {49}},\ \bibinfo {pages} {1567} (\bibinfo
  {year} {1994})}\BibitemShut {NoStop}%
\bibitem [{\citenamefont {Simon}(2000)}]{SimonPRL2000}%
  \BibitemOpen
  \bibfield  {author} {\bibinfo {author} {\bibfnamefont {R.}~\bibnamefont
  {Simon}},\ }\bibfield  {title} {\enquote {\bibinfo {title} {Peres-horodecki
  separability criterion for continuous variable systems},}\ }\href {\doibase
  10.1103/PhysRevLett.84.2726} {\bibfield  {journal} {\bibinfo  {journal}
  {Phys. Rev. Lett.}\ }\textbf {\bibinfo {volume} {84}},\ \bibinfo {pages}
  {2726--2729} (\bibinfo {year} {2000})}\BibitemShut {NoStop}%
\bibitem [{\citenamefont {Agarwal}\ and\ \citenamefont
  {Huang}(2010)}]{agarwal2010electromagnetically}%
  \BibitemOpen
  \bibfield  {author} {\bibinfo {author} {\bibfnamefont {GS}~\bibnamefont
  {Agarwal}}\ and\ \bibinfo {author} {\bibfnamefont {Sumei}\ \bibnamefont
  {Huang}},\ }\bibfield  {title} {\enquote {\bibinfo {title}
  {Electromagnetically induced transparency in mechanical effects of light},}\
  }\href@noop {} {\bibfield  {journal} {\bibinfo  {journal} {Physical Review
  A}\ }\textbf {\bibinfo {volume} {81}},\ \bibinfo {pages} {041803} (\bibinfo
  {year} {2010})}\BibitemShut {NoStop}%
\bibitem [{\citenamefont {Tarhan}\ \emph {et~al.}(2013)\citenamefont {Tarhan},
  \citenamefont {Huang},\ and\ \citenamefont
  {M{\"u}stecapl{\i}o{\u{g}}lu}}]{tarhan2013superluminal}%
  \BibitemOpen
  \bibfield  {author} {\bibinfo {author} {\bibfnamefont {Devrim}\ \bibnamefont
  {Tarhan}}, \bibinfo {author} {\bibfnamefont {Sumei}\ \bibnamefont {Huang}}, \
  and\ \bibinfo {author} {\bibfnamefont {{\"O}zg{\"u}r~E}\ \bibnamefont
  {M{\"u}stecapl{\i}o{\u{g}}lu}},\ }\bibfield  {title} {\enquote {\bibinfo
  {title} {Superluminal and ultraslow light propagation in optomechanical
  systems},}\ }\href@noop {} {\bibfield  {journal} {\bibinfo  {journal}
  {Physical Review A}\ }\textbf {\bibinfo {volume} {87}},\ \bibinfo {pages}
  {013824} (\bibinfo {year} {2013})}\BibitemShut {NoStop}%
\bibitem [{\citenamefont {Tasgin}(2015)}]{tasgin2015measure}%
  \BibitemOpen
  \bibfield  {author} {\bibinfo {author} {\bibfnamefont {Mehmet~Emre}\
  \bibnamefont {Tasgin}},\ }\bibfield  {title} {\enquote {\bibinfo {title}
  {Single-mode nonclassicality measure from simon-peres-horodecki criterion},}\
  }\href@noop {} {\bibfield  {journal} {\bibinfo  {journal} {arXiv preprint
  arXiv:1502.00992}\ } (\bibinfo {year} {2015})}\BibitemShut {NoStop}%
\bibitem [{\citenamefont {Ge}\ \emph {et~al.}(2015)\citenamefont {Ge},
  \citenamefont {Tasgin},\ and\ \citenamefont {Zubairy}}]{ge2015conservation}%
  \BibitemOpen
  \bibfield  {author} {\bibinfo {author} {\bibfnamefont {Wenchao}\ \bibnamefont
  {Ge}}, \bibinfo {author} {\bibfnamefont {Mehmet~Emre}\ \bibnamefont
  {Tasgin}}, \ and\ \bibinfo {author} {\bibfnamefont {M~Suhail}\ \bibnamefont
  {Zubairy}},\ }\bibfield  {title} {\enquote {\bibinfo {title} {Conservation
  relation of nonclassicality and entanglement for gaussian states in a beam
  splitter},}\ }\href@noop {} {\bibfield  {journal} {\bibinfo  {journal}
  {Physical Review A}\ }\textbf {\bibinfo {volume} {92}},\ \bibinfo {pages}
  {052328} (\bibinfo {year} {2015})}\BibitemShut {NoStop}%
\bibitem [{\citenamefont {Scully}\ and\ \citenamefont
  {Zubairy}(1997)}]{ScullyZubairyBook}%
  \BibitemOpen
  \bibfield  {author} {\bibinfo {author} {\bibfnamefont {M.~O.}\ \bibnamefont
  {Scully}}\ and\ \bibinfo {author} {\bibfnamefont {M.~S.}\ \bibnamefont
  {Zubairy}},\ }\href@noop {} {\emph {\bibinfo {title} {Quantum Optics}}}\
  (\bibinfo  {publisher} {Cambridge University Press},\ \bibinfo {address} {New
  York},\ \bibinfo {year} {1997})\BibitemShut {NoStop}%
\bibitem [{\citenamefont {Arkhipov}\ \emph
  {et~al.}(2016{\natexlab{a}})\citenamefont {Arkhipov}, \citenamefont
  {Pe{\v{r}}ina~Jr}, \citenamefont {Svozil{\'\i}k},\ and\ \citenamefont
  {Miranowicz}}]{arkhipov2016nonclassicality}%
  \BibitemOpen
  \bibfield  {author} {\bibinfo {author} {\bibfnamefont {Ievgen~I}\
  \bibnamefont {Arkhipov}}, \bibinfo {author} {\bibfnamefont {Jan}\
  \bibnamefont {Pe{\v{r}}ina~Jr}}, \bibinfo {author} {\bibfnamefont
  {Ji{\v{r}}{\'\i}}\ \bibnamefont {Svozil{\'\i}k}}, \ and\ \bibinfo {author}
  {\bibfnamefont {Adam}\ \bibnamefont {Miranowicz}},\ }\bibfield  {title}
  {\enquote {\bibinfo {title} {Nonclassicality invariant of general two-mode
  gaussian states},}\ }\href@noop {} {\bibfield  {journal} {\bibinfo  {journal}
  {Scientific reports}\ }\textbf {\bibinfo {volume} {6}},\ \bibinfo {pages}
  {26523} (\bibinfo {year} {2016}{\natexlab{a}})}\BibitemShut {NoStop}%
\bibitem [{\citenamefont {Arkhipov}\ \emph
  {et~al.}(2016{\natexlab{b}})\citenamefont {Arkhipov}, \citenamefont
  {Pe{\v{r}}ina~Jr}, \citenamefont {Pe{\v{r}}ina},\ and\ \citenamefont
  {Miranowicz}}]{arkhipov2016interplay}%
  \BibitemOpen
  \bibfield  {author} {\bibinfo {author} {\bibfnamefont {Ievgen~I}\
  \bibnamefont {Arkhipov}}, \bibinfo {author} {\bibfnamefont {Jan}\
  \bibnamefont {Pe{\v{r}}ina~Jr}}, \bibinfo {author} {\bibfnamefont {Jan}\
  \bibnamefont {Pe{\v{r}}ina}}, \ and\ \bibinfo {author} {\bibfnamefont {Adam}\
  \bibnamefont {Miranowicz}},\ }\bibfield  {title} {\enquote {\bibinfo {title}
  {Interplay of nonclassicality and entanglement of two-mode gaussian fields
  generated in optical parametric processes},}\ }\href@noop {} {\bibfield
  {journal} {\bibinfo  {journal} {Physical Review A}\ }\textbf {\bibinfo
  {volume} {94}},\ \bibinfo {pages} {013807} (\bibinfo {year}
  {2016}{\natexlab{b}})}\BibitemShut {NoStop}%
\bibitem [{\citenamefont {{\v{C}}ernoch}\ \emph {et~al.}(2018)\citenamefont
  {{\v{C}}ernoch}, \citenamefont {Bartkiewicz}, \citenamefont {Lemr},\ and\
  \citenamefont {Soubusta}}]{vcernoch2018experimental}%
  \BibitemOpen
  \bibfield  {author} {\bibinfo {author} {\bibfnamefont {Anton{\'\i}n}\
  \bibnamefont {{\v{C}}ernoch}}, \bibinfo {author} {\bibfnamefont {Karol}\
  \bibnamefont {Bartkiewicz}}, \bibinfo {author} {\bibfnamefont {Karel}\
  \bibnamefont {Lemr}}, \ and\ \bibinfo {author} {\bibfnamefont {Jan}\
  \bibnamefont {Soubusta}},\ }\bibfield  {title} {\enquote {\bibinfo {title}
  {Experimental tests of coherence and entanglement conservation under unitary
  evolutions},}\ }\href@noop {} {\bibfield  {journal} {\bibinfo  {journal}
  {Physical Review A}\ }\textbf {\bibinfo {volume} {97}},\ \bibinfo {pages}
  {042305} (\bibinfo {year} {2018})}\BibitemShut {NoStop}%
\bibitem [{\citenamefont {Roelli}\ \emph {et~al.}(2016)\citenamefont {Roelli},
  \citenamefont {Galland}, \citenamefont {Piro},\ and\ \citenamefont
  {Kippenberg}}]{SERSoptomechanicsNatureNano2016}%
  \BibitemOpen
  \bibfield  {author} {\bibinfo {author} {\bibfnamefont {Philippe}\
  \bibnamefont {Roelli}}, \bibinfo {author} {\bibfnamefont {Christophe}\
  \bibnamefont {Galland}}, \bibinfo {author} {\bibfnamefont {Nicolas}\
  \bibnamefont {Piro}}, \ and\ \bibinfo {author} {\bibfnamefont {Tobias~J}\
  \bibnamefont {Kippenberg}},\ }\bibfield  {title} {\enquote {\bibinfo {title}
  {Molecular cavity optomechanics as a theory of plasmon-enhanced raman
  scattering},}\ }\href@noop {} {\bibfield  {journal} {\bibinfo  {journal}
  {Nature nanotechnology}\ }\textbf {\bibinfo {volume} {11}},\ \bibinfo {pages}
  {164} (\bibinfo {year} {2016})}\BibitemShut {NoStop}%
\bibitem [{\citenamefont {Garrett}\ \emph {et~al.}(1997)\citenamefont
  {Garrett}, \citenamefont {Rojo}, \citenamefont {Sood}, \citenamefont
  {Whitaker},\ and\ \citenamefont {Merlin}}]{phonon_squeezing_Science_1997}%
  \BibitemOpen
  \bibfield  {author} {\bibinfo {author} {\bibfnamefont {GA}~\bibnamefont
  {Garrett}}, \bibinfo {author} {\bibfnamefont {AG}~\bibnamefont {Rojo}},
  \bibinfo {author} {\bibfnamefont {AK}~\bibnamefont {Sood}}, \bibinfo {author}
  {\bibfnamefont {JF}~\bibnamefont {Whitaker}}, \ and\ \bibinfo {author}
  {\bibfnamefont {R}~\bibnamefont {Merlin}},\ }\bibfield  {title} {\enquote
  {\bibinfo {title} {Vacuum squeezing of solids: macroscopic quantum states
  driven by light pulses},}\ }\href@noop {} {\bibfield  {journal} {\bibinfo
  {journal} {Science}\ }\textbf {\bibinfo {volume} {275}},\ \bibinfo {pages}
  {1638--1640} (\bibinfo {year} {1997})}\BibitemShut {NoStop}%
\bibitem [{\citenamefont {Tasgin}\ \emph {et~al.}(2019)\citenamefont {Tasgin},
  \citenamefont {Gunay},\ and\ \citenamefont
  {Zubairy}}]{tasgin2019Wavepackets}%
  \BibitemOpen
  \bibfield  {author} {\bibinfo {author} {\bibfnamefont {Mehmet~Emre}\
  \bibnamefont {Tasgin}}, \bibinfo {author} {\bibfnamefont {Mehmet}\
  \bibnamefont {Gunay}}, \ and\ \bibinfo {author} {\bibfnamefont {M~Suhail}\
  \bibnamefont {Zubairy}},\ }\bibfield  {title} {\enquote {\bibinfo {title}
  {Nonclassicality and entanglement for wavepackets},}\ }\href@noop {}
  {\bibfield  {journal} {\bibinfo  {journal} {arXiv preprint arXiv:1904.13149}\
  } (\bibinfo {year} {2019})}\BibitemShut {NoStop}%
\end{thebibliography}%

\newpage

\begin{figure}
\begin{center}
\includegraphics[width=0.8\textwidth]{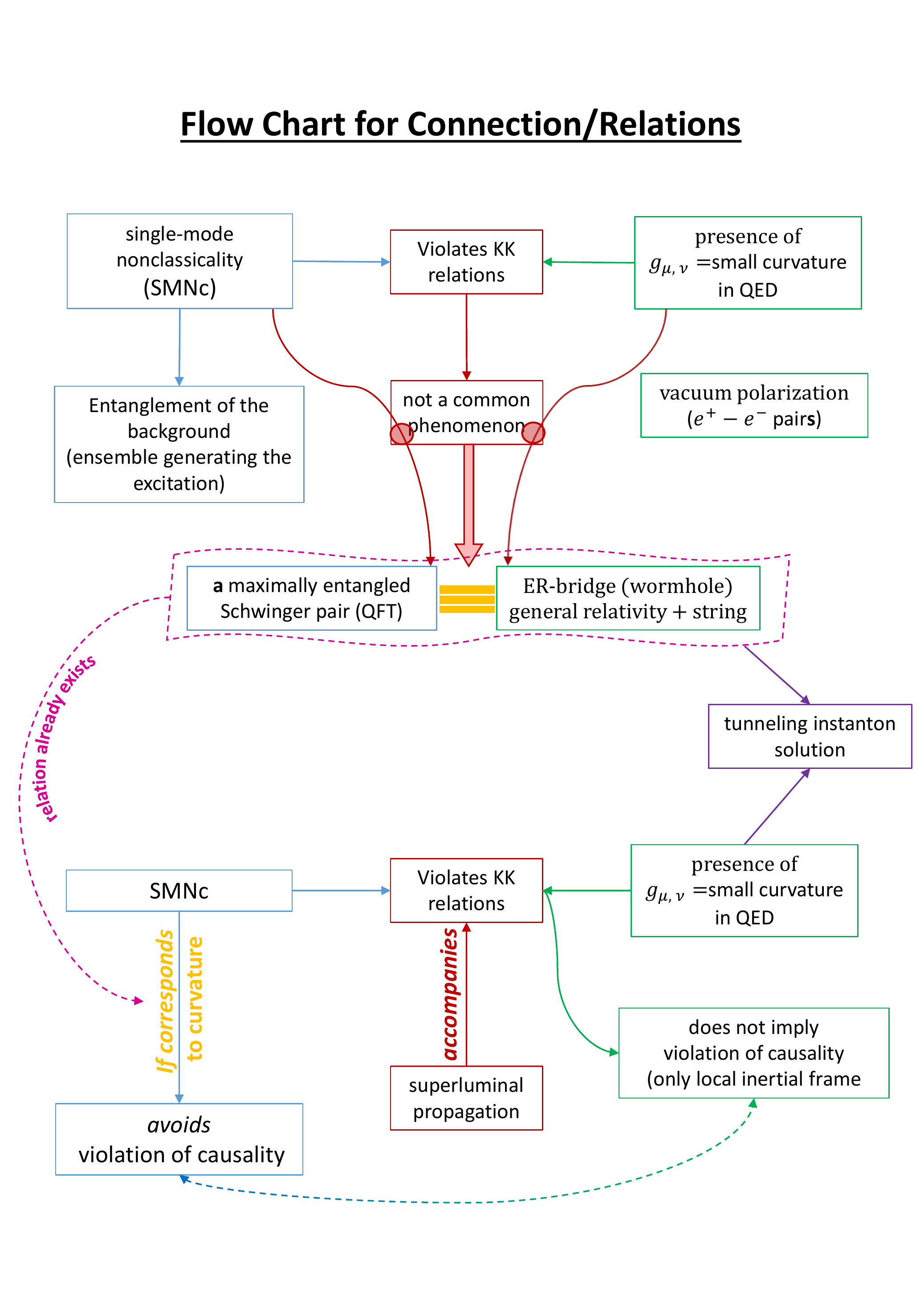}
\end{center}
\end{figure}

  
\end{document}